\documentclass{ar-1col}
\jname{Xxxx. Xxx. Xxx. Xxx.}
\jvol{AA}
\jyear{YYYY}
\doi{10.1146/((please add article doi))}

\begin{document}

\markboth{Arpita Paul, Turan Birol}{Applications of DFT+DMFT}

\title{Applications of DFT+DMFT in Materials Science}

\author{Arpita Paul,$^1$ Turan Birol$^{1,*}$
\affil{$^1$Department of Chemical Engineering and Materials Science, University of Minnesota, Minneapolis, Minnesota 55455, USA}
\affil{$^*$e-mail: tbirol@umn.edu}}

\begin{abstract}
First principles methods can provide insight into materials that otherwise is impossible to acquire. 
Density Functional Theory (DFT) has been the first principles method of choice for numerous applications, but 
it falls short of predicting the properties of correlated materials. 
First principles Density Functional Theory + Dynamical Mean Field Theory (DFT+DMFT) is a 
powerful tool that can address these shortcomings of DFT when applied to correlated metals. In this brief 
review, which is aimed at non-experts, we review the basics and some applications of DFT+DMFT.  
\end{abstract}

\begin{keywords}
first principles calculations, density functional theory, dynamical mean field theory, correlated materials
\end{keywords}
\maketitle

\tableofcontents

\section{INTRODUCTION}
As early as the first quarter of 20th century, the laws of quantum mechanics were almost completely known,
and it was realized that understanding the properties of crystalline matter was in principle a problem of
solving the Schroedinger equation for the electrons. In 1929 Paul Dirac, one of the founding fathers
of quantum mechanics, published an article titled
\textit{Quantum Mechanics of Many-Electron Systems},
in which he famously claimed that \cite{dirac1929}
\begin{extract}
``The general theory of quantum mechanics is now almost complete, the
imperfections that still remain being in connection with the exact fitting in
of the theory with relativity ideas. (...) The underlying physical laws necessary
for the mathematical theory of a large part of physics and the whole of chemistry
are thus completely known, and the difficulty is only that the exact application
of these laws leads to equations much too complicated to be soluble. 
It therefore
becomes desirable that approximate practical methods of applying quantum
mechanics should be developed, which can lead to an explanation of the main
features of complex atomic systems without too much computation.''
\end{extract}
A plethora of numerical methods were developed in the course of the past century 
with the aim of achieving an accurate enough solution to the Schroedinger equation for electrons      
in molecules and solids \cite{szabo2012, martin2004}. Those different methods have achieved different levels of success.
For example, the Thomas-Fermi Theory, which predates even the quote above, is good for qualitatively explaining 
total energies of atoms, but cannot predict any chemical bonding \cite{kohn1999}.
The Hartree-Fock approach, on the other hand, is capable of reproducing various chemical phenomena but only at the  
cost of a much higher computational cost, and fails to capture electronic states that cannot be represented as
a single Slater determinant.

The workhorse method for solid state materials physics is the Density Functional Theory (DFT).\footnote{The idea of using density as the 
basic variable and forming a theoretical framework that relies on functionals of density is a very general one,  
and it is possible to construct different DFT approaches. However, the dominant convention in the electronic 
structure community is to refer to a particular DFT framework, 
the Kohn-Sham DFT combined with the Local Density Approximation (LDA) or Generalized Gradient 
Approximations (GGA), as \textit{the} DFT, and throughout this paper, we follow this convention as well.} 
The state of the art DFT is extremely successful, and reproducible \cite{lejaeghere2016}; however it also 
has fundamental shortcomings, such as the absence of dynamic electronic correlations which are important in 
Mott insulators.   
The Dynamical Mean Field Theory (DMFT) is an approach that was originally developed to solve the Hubbard model, 
and it was interfaced with DFT to become first principles DFT+DMFT soon after \cite{anisimov1997}. 
Its successes include, but are not limited to, 
the prediction of correct electronic structure for Mott insulators and correlated metals \cite{kotliar2004}. 
With more and more applications 
of first principles DFT+DMFT to novel materials systems, this method is no longer used exclusively by the physicists focused 
on the correlated electronic phases but is now becoming a widely used tool for the materials science community 
as well \cite{kent2018}.  

The aim of this brief review article is to use some successful applications of the DFT+DMFT method to demonstrate 
its capabilities to non-experts. In this respect, it is neither a complete review of the intricacies of this method, 
nor is it even a nearly complete list 
of applications of DFT+DMFT. It is rather a short introduction for experimentalists and theorists focused 
on other approaches, and we refer the reader to many excellent review articles on the fundamentals and applications 
of this method for further information \cite{kotliar2006, held2007, basov2011, georges1996, kotliar2004, martin2016}.  

This article is organized as follows: In Section \ref{sec:background}, we give a very brief background on the DFT and DMFT. 
In Section \ref{sec:mass}, 
we demonstrate how DFT+DMFT can correct the spectra and reproduce the correlation induced mass enhancement in correlated 
metals. Section \ref{sec:structure} demonstrates the DMFT corrections to the crystal structures and phonon spectra 
of correlated materials. Section \ref{sec:nonlocal} provides examples of the extensions of DMFT 
to include nonlocal correlations. We conclude by Section \ref{sec:summary}, a summary.

\section{DFT, DMFT, and DFT+DMFT}
\label{sec:background}
\subsection{Density Functional Theory}
The starting point of DFT is the observation that the many-electron wavefunction $\Psi$ is a prohibitively expensive
function to numerically work with. For $N$ electrons, $\Psi$ is a function of $3N$ variables such as the cartesian coordinates
$x_1$, $y_1$, $z_1$, $x_2$, ..., $z_N$. The number of bits required to numerically store such a function scales exponentially with 
the number of electrons $N$ \cite{kohn1999}, and becomes larger than the number of protons in the observable universe 
for even relatively humble molecules. This limits the applicability of numerical approaches that rely on brute force calculation of the 
wavefunction. 

One way to circumvent this exponential barrier is to use the electron density $\rho$, instead of the many-electron wavefunction 
$\Psi$. Being only a function of three spatial coordinates, $\rho(x,y,z)$ does not suffer from the same exponential scaling. 
However, being a real function in a much fewer dimensional space, $\rho$ might seem to hold much less information than $\Psi$. 
Nevertheless, the Hohenberg-Kohn Theorem of 1964 \cite{hohenberg1964} states that $\rho$ implicitly carries all the 
necessary information about the groundstate properties. Citing \cite{kohn1999}: 
\begin{extract}
``The groundstate density of a bound system of interacting electrons in some external potential $V$ determines this potential uniquely.'' 
\end{extract}
In the context of a crystal, the external potential $V$ is the electrostatic potential of the ion cores. Since $\rho$ determines $V$, it 
also determines the full electronic Hamiltonian. As a result, the groundstate density $\rho$ carries all the information about the 
physical system. In other words, there is in principle a functional of density $\rho$ for any physical observable. However, in practice, 
these functionals are not known, and only approximate DFT calculations are performed. 

The Kohn-Sham DFT relies on solving a noninteracting problem instead of the interacting one by defining an effective Coulomb potential 
$V_C$ that stems from both the external (ionic) potential $V$, and the electrons
\begin{equation}
V_C(\vec{r})=V(\vec{r})+\int \frac{\rho(\vec{r}')}{|\vec{r} -\vec{r}'|}d\vec{r}' 
\end{equation}
and an \textit{exchange correlation functional} $V_{xc}$. 
The ground state density $\rho(\vec{r})$ is the solution of the self consistent equations 
\begin{equation}
\left(-\frac{1}{2}\nabla^2 +V_C(\vec{r}) + V_{xc}(\vec{r}) -E_i\right)\psi_i(\vec{r})=0
\end{equation}
and
\begin{equation}
\rho(\vec{r})=\sum_i|\psi_i(\vec{r})|^2 
\end{equation}
Here, $\psi_i$ are the wavefunctions for the Kohn-Sham quasiparticles, which are noninteracting. This simplifies the 
$N$ electron problem to $N$ one-electron problems. 
This form of the equations is exact, and the Kohn-Sham eigenvalues $E_i$ can be used to calculate the total ground state energy 
of the system. However, the form of the exchange correlation functional $V_{xc}$ is not known. Even though it is often written 
as a local function, the value of $V_{xc}$ at $\vec{r}$ depends on the electron density distribution in the whole space. While its overall 
magnitude is small compared to the Hartree potential $V_C$, the errors in approximating the exchange-correlation functional can 
lead to qualitative errors. It is the shortcomings of the present approximations to $V_{xc}$ that makes DFT unreliable in certain 
types of materials systems. 

The simplest approximation to the exchange correlation functional is to assume that it is a local function that only depends on the 
magnitude of the electron density at point $\vec{r}$, and it is equal to that of a homogeneous electron gas with equal density. 
This leads to the so-called local density approximation (LDA). The value of the exchange-correlation energy of the homogeneous electron 
gas can be calculated numerically with arbitrary precision, as has been done by Ceperley and Alder in 1980 using a 
Monte Carlo approach \cite{ceperley1980}. LDA is expected to work well in the limit that the electron density changes slowly with $\vec{r}$, 
however, what quantity defines the slow change is far from obvious. Nevertheless, DFT with the LDA works surprisingly 
well for a wide range of systems, including many molecular systems, and crystalline systems such as band insulators or uncorrelated metals. 
Various Generalized Gradient Approximations that take into account the derivative of the electron density are commonly used 
to approximate $V_{xc}$ as well \cite{perdew1991}. While GGAs provide better quantitative results for certain quantities such as the lattice 
constants or binding energies, they suffer from the same fundamental shortcomings of the LDA.

\subsection{Dynamical Mean Field Theory}
The large spatial extent of electrons in the $s$ and $p$ orbitals of atoms and the broad energy bands formed by these orbitals in solids 
facilitate the screening of the intra-atomic Coulomb interaction that is effectively felt by these electrons. 
These electrons are highly itinerant and their large kinetic energies dominate over the Coulomb interaction.  
Usually, a static mean field approximation is suitable to describe $s$ and $p$ electrons because their wavefunctions are 
not strongly correlated, and can be expressed by a single Slater determinant. Band theory, which relies on 
an independent electron approximation, treats electrons as Bloch 
waves and is a very successful approach in theoretically understanding the physical properties of simple metals, 
semiconductors, and band insulators. Success of first principles DFT when applied to these materials, 
in part, relies on this observation. 

In contrast, the Coulomb interaction between electrons in the $d$ or $f$ orbitals is often stronger, as electrons in these 
orbitals with a smaller radial extent and larger number of angular nodes lead to narrow energy bands that lead to weaker 
screening. Strongly correlated electronic states that cannot be represented in a single 
Slater determinant emerge often in these systems, and the band theory fails.\footnote{In principle, since the many-body electronic 
wavefunction is antisymmetric, any electronic state is correlated. However, a correlated electronic state is usually defined as 
one that cannot be represented by a single Slater determinant, and correlation effects are defined as those that stem from this fact 
\cite{giuliani2005}. Often, effects not captured by DFT are also referred to as correlation effects, though this usage can be misleading.}

%
There are several emergent phenomena and phases induced by the correlations of $d$ and $f$ electrons, and couplings between
multiple competing degrees of freedom such as spin, charge, orbital, and lattice \cite{vollhardt2012,kotliar2004}. 
Examples of such phenomena, which are beyond a simple band theory, include the high temperature superconductivity 
in cuprates \cite{anderson2013}, the colossal magnetoresistance in manganites \cite{dagotto2013}, Mott metal-insulator 
transitions \cite{imada1998}, and the mass enhancement of electrons in heavy Fermion systems. 
Because of this competition, strongly correlated electron systems are often extremely sensitive to external 
perturbations such as temperature, pressure, doping and magnetic field \cite{imada1998}, which renders them both experimentally 
and theoretically interesting materials to study.

In the limit of strong interactions, electrons become highly localized on atomic sites, and eventually the solid becomes a Mott 
insulator. Electrons in the Mott insulating state are better described by an atomic-like theory defined in real space, rather than 
by band theory in reciprocal space; and as a result the wavevector $\vec{k}$ is no longer a good quantum number.
The failure of band theory was first observed in insulating transition metal compounds like MnO and NiO by predicting these 
to be metallic in the absence of long range magnetic ordering \cite{mott1949, imada1998,Terakura-1984}.
In the regime of intermediate interactions, on the other hand, electrons are not fully localized, and can display features of both 
Bloch-like bands and localization (such as the quasiparticle bands not being sharp, and emergence of upper and lower Hubbard bands).
In this correlated metallic regime, it is necessary to consider both natures of electrons, and use a method that takes advantage of both 
real and reciprocal spaces. 
Both band theory and an atomic-like theory fail to explain this particular behavior of electrons alone. 

Early work on understanding the correlated electronic structure was focused on the Hubbard model 
\cite{Gutzwiller-1963,hubbard1963,Kanamori-1963}, which includes a local Coulomb repulsion $U$ between electrons: 
\begin{equation} \label{DMFT_eq1}
\hat{H}=\sum_{ij,\sigma} t_{ij}c^{\dagger}_{i\sigma}c_{j\sigma}+U\sum_{i}n_{i\uparrow}n_{i\downarrow}.
\end{equation}
Here, $c^{\dagger}_{i\sigma}$ and $c_{i\sigma}$ are the creation and annihilation operators associated with the electron with 
spin $\sigma$ at $i^{th}$ lattice site, and $t_{ij}$ is the inter-site hopping amplitude. In the $t \gg U$ limit, 
electrons are itinerant and the band theory works well. 
In the $U \gg t$ limit double occupation of a site is energetically unfavorable, and the system becomes a Mott insulator at half 
filling. (The large electron scattering at the Fermi level introduces a gap separating the upper and lower Hubbard bands.) 
A static mean field theory like Hartree-Fock or DFT cannot capture the dynamic (frequency dependent) correlations that emerges 
from the strong interactions and scattering between the electrons, and thus cannot predict this Mott insulating state. 
Dynamical correlations are important in the $t\sim U$ regime as well. In this regime, the width of the quasiparticle band 
is renormalized, the quasiparticles attain a finite life time (the bands become partially incoherent), and upper and lower 
Hubbard bands emerge.

The one-band Hubbard model is exactly solvable in one dimension \cite{Lieb-1968} but 
not in 2 or 3 dimensions \cite{Lieb-1968,pavarini2014dmft,vollhardt2012}. 
As the number of lattice sites increases, the Hilbert 
space expands exponentially and the many body problem becomes computationally intractable even in modern day supercomputers 
\cite{vollhardt2012}. 
As early as the 1980s, it was realized that the Hubbard model is more tractable in the limit of infinite dimensions \cite{Metzner-1989} 
where the electronic self energy becomes independent of the wavevector $\vec{k}$ \cite{mullerhartmann1989}. 
Subsequently, Georges and Kotliar formulated the idea of mapping the Hubbard model in the infinite dimensional limit into a 
self consistent single site quantum impurity model, and hence laid the foundations of the dynamical mean field theory (DMFT) 
approach \cite{georges1992,georges1996}. This non-perturbative approach has directed towards significant advancement in 
understanding strongly correlated systems 
\cite{vollhardt2012,kotliar2004,georges1996,georges2004,vollhardt2010,kotliar2006,georges1992,Jarrell-1992}. 
DMFT can be considered as an analogue of a classical mean field theory for a ferromagnetic system: The classical and static mean field 
theory for the magnetic system introduces a magnetic field that is induced by the average magnetization of the whole crystal acting on 
each magnetic atom. The magnetic configuration of the atom and the mean magnetic field acting on it are determined self consistently. 
In a similar vein, DMFT replaces the many-body system with a single impurity atom which is embedded in a bath of uncorrelated 
electrons, and determines the hybridization between the impurity and the bath self consistently. 
A major strength of this approach is the possibility to treat the quasiparticle and Hubbard bands equally\cite{kotliar2006,kotliar2004}.

In DMFT, the many-body problem is mapped onto the well known Anderson impurity model (AIM), 
which is often used to model magnetic impurities embedded in metals \cite{Anderson-1961,kotliar2004}: 
\begin{equation}
\mathcal{H}_{AIM}=\mathcal{H}_{atom}+\sum_{\nu,\sigma}\epsilon_{\nu}^{bath}n_{\nu,\sigma}^{bath}+\sum_{\nu,\sigma}(V_{\nu}c_{0,\sigma}^{\dagger}a_{\nu,\sigma}^{bath}+H.C.).
\end{equation}
In this Hamiltonian, $\mathcal{H}_{atom}$ represents the energy associated with the atomic degrees of freedom at the impurity site, 
$\epsilon_{\nu}^{bath}$ are the energy levels of the bath of noninteracting electrons, $n_{\sigma}$=$c_{\sigma}^{\dagger}c_{\sigma}$ 
is the density of electrons with spin $\sigma$, and $V_{\nu}$ is the probability amplitude of an electron being exchanged between the 
impurity and the bath. The frequency dependent hybridization function $\Delta(\omega)$ is defined by $V_{\nu}$ as \cite{kotliar2004}
\begin{equation}
\Delta(\omega)=\sum_{\nu}\frac{|V_{\nu}|^{2}}{\omega-\epsilon_{\nu}^{bath}}.
\end{equation}
The dynamic quantity $\Delta$($\omega$) serves as the dynamical mean field in DMFT, and is the analogue of the Weiss mean field in 
classical mean field theory of magnetism.
The hybridization function determines the ability of electron to hop in and out of the impurity site. The electrons become 
localized or itinerant for small and large values of $\Delta$ respectively. 

\begin{figure}[h]
\includegraphics[width=3.5in]{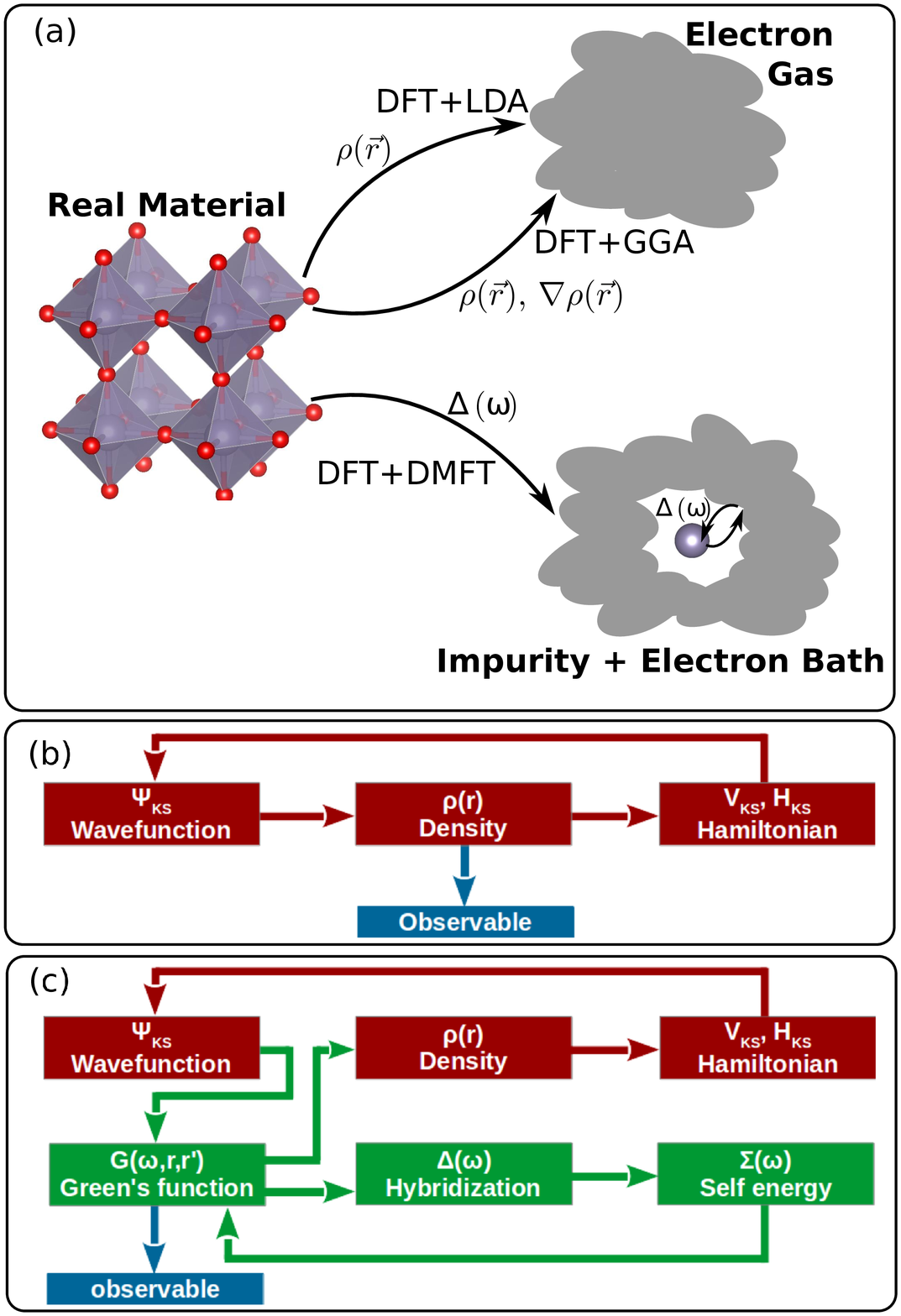}
\caption{(a) The LDA approximation maps the electronic system to a homogeneous electron gas. In contrast, DMFT 
maps the system to an impurity that interacts with an electron bath via the hybridization function $\Delta(\omega)$. 
(b) Typical DFT implementations calculate the Kohn-Sham wavefunctions, electron density, and the Kohn-Sham Hamiltonian 
iteratively until self consistency is reached. The observables of interest are in principle determined by the electron 
density. (In practice, the Kohn-Sham wavefunctions are also used.) (c) Typical DFT+DMFT implementation, in addition to performing the 
DFT iterations, involves calculating the Green's function, Hybridization $\Delta(\omega)$, and the electronic 
self energy $\Sigma(\omega)$.}
\label{fig:methods}
\end{figure}

The DMFT method can be compared with DFT as shown in Fig. \ref{fig:methods}a \cite{haule2018}. 
DFT at the level of LDA approximation models the 
electrons in the real material by non-interacting Kohn-Sham quasiparticles. The exchange correlation energy is calculated 
by using a model of a homogeneous electron gas with the same density as $\rho(\vec{r})$. DFT+GGA is uses derivatives 
of the charge density $\rho(\vec{r})$ as well, but the model that the system is mapped onto is nevertheless featureless. 
DMFT is significantly more involved, and considers an impurity atom, such as a transition metal ion with, with all its internal 
degrees of freedom. The interactions of the impurity atom with an uncorrelated bath of electrons are taken into 
account via the hybridization function $\Delta(\omega)$. The 
information about the real material that is retained in the AIM is this hybridization function, which is 
considerably larger than the information carried by the local electron density alone. 

\subsection{First Principles DFT+DMFT}
DMFT applied to the Hubbard Model led to great advancements in the understanding of the behaviour of the strongly correlated 
systems, and the nature of phenomena such as the Mott transition. However, DMFT is blind to chemistry: in order to apply 
DMFT to real materials, it needs to be interfaced with a first principles method like DFT. This is achieved by performing 
a DFT calculation, and then defining the atoms with $d$ or $f$ electrons as the impurities in the AIM. 
First attempts to perform DFT+DMFT followed the procedure of building a tight binding model using the first principles calculations, and 
using the DMFT approach to solve the tight binding model. The most common approach to obtain the first principles tight binding model 
is to use the maximally localized Wannier orbitals \cite{marzari2012}. 
The shortcoming of this `one-shot' approach, however, is that the orbitals that define 
the DMFT problem are built using a method that does not take into effect the electronic correlations (DFT). 
It is possible to include the effect of the correlations as given by DMFT on the charge density by repeating the 
DFT calculation while taking into account the charge density updated by DMFT \cite{park2014energy}, 
and almost all modern implementations employ 
such a self-consistent DFT+DMFT loop. (This, of course, comes with an increased computational cost.)
There are also projector-based approaches, which do not suffer from the errors introduced by the Wannierization 
procedure, such as the possible change of the extent of the orbitals which affects the effective on-site interaction 
strength \cite{haule2010}.

In a typical DFT calculation, the Kohn-Sham wavefunctions, electron density, and the Kohn-Sham Hamiltonian are determined 
self-consistently, as shown in Fig. \ref{fig:methods}b. This DFT loop also exists in its entirety in the flow-chart for a 
DFT+DMFT calculation, a simplified version of which is shown in Fig. \ref{fig:methods}c. The Kohn-Sham wavefunctions are 
used to calculate the Green's function, which in turn determines the impurity hybridization function $\Delta(\omega)$. Solving 
the impurity problem, computationally the most expensive step, gives the impurity self energy $\Sigma(\omega)$. The self energy 
updates the Green's function, which can be fed back to the DFT loop to update the charge density. A typical calculation involves going over 
both the DFT and DFT+DMFT loops many times to reach self consistency. 

The DFT+DMFT approach is now well developed and tested enough to have predictive capabilities, and is used extensively on 
materials science problems \cite{kent2018, adler2018}. However, like any other method, it still has room for technical improvements. 
One of the most important problems of DFT+DMFT method is considered to be the double-counting (DC) issue, 
which arises in any electronic structure 
method that incorporates additional interaction terms onto DFT \cite{DFTU:LDAUTYPE1_1,FLL_DC_Karolak-2010}. As DFT includes a certain 
part of the static correlations of electrons through the exchange-correlation functionals (LDA or GGA), it is necessary 
to subtract the part of energy that is accounted for twice. This double counted part of the correlation energy shifts the energies of 
correlated states with respect to the uncorrelated ones, and can give rise to errors in the final electronic structure:  
for example, the $p-d$ charge transfer energy of transition metal oxides often depend on the choice of 
DC scheme \cite{FLL_DC_Marianetti_2014}. 
Historically, simple expressions for the DC energy were borrowed from the DFT+U literature, and this 
problem was one of the most common reasons used to claim that DFT+DMFT is not a true first principles method \cite{haule2018}. 
The two prevalent approaches were the ``Fully Localized Limit'' \cite{Sawatzky-1994} and the ``Around Mean Field'' \cite{DFTU:LDAUTYPE1_1} 
formulas. However, these formulas often need to be `tuned' with only a posteriori justification \cite{park2014}. 
A recent development on the DC problem is the derivation of an exact DC expression 
using a continuum representation of DMFT \cite{Haule-2015}. This approach takes into account the nonsphericity of the impurity, and 
has so far produced good agreement with the experiment \cite{haule2018, Kristanovski-2018}, possibly concluding the discussion over 
the different DC approaches. 

Another area where there is need for new developments is the numerical impurity solvers, which 
solve the AIM to calculate the self energy $\Sigma(\omega)$ from the hybridization $\Delta(\omega)$ \cite{kent2018}. The state of the art 
is the continuous time quantum Monte Carlo methods introduced for this purpose \cite{werner2006, haule2007, gull2011} which are 
in principle exact, and efficiently parallelizable over a large number of processors. 
However, their applicability is strongly limited by the computational cost for 
large number of impurity orbitals, and reduced Monte Carlo noise. New algorithmic developments, such as the `lazy skip lists' 
are introduced to reduce the computational needs \cite{semon2014}, but applications of state of the art DMFT on problems with 
large impurities, such as certain $f$ electron compounds, and clusters of transition metals, are still hardly within reach except for 
large scale supercomputers.

\section{MASS RENORMALIZATION IN CORRELATED METALS}
\label{sec:mass}
How the inclusion of electronic correlations at the DMFT level changes the DFT band structure can be illustrated 
by considering the electronic Green's function $G(\vec{k},\omega)$, which is the fourier transform of the electron 
propagation amplitude. For a one-band system: 
\begin{equation}
G(\vec{k},\omega)=\frac{1}{\omega - E(\vec{k})-\Sigma(\omega, \vec{k})}
\end{equation}
Here, $E(\vec{k})$ is the energy of the band at $\vec{k}$ according to DFT, and $\Sigma(\vec{k}, \omega)$ is the self energy 
obtained from the DMFT calculation. 
If $\Sigma\!=\!0$, 
the Green's function has poles at $\omega=E(\vec{k})$, and the spectral function 
\begin{equation}
A(\vec{k},\omega)=-\frac{1}{\pi} Im(G(\vec{k},\omega))
\end{equation}
consists of a Dirac delta for each $\vec{k}$ at $\omega=E(\vec{k})$. Hence, the spectral function is equivalent to the 
DFT band structure. 

The self energy $\Sigma(\omega, \vec{k})$ is in general a complex function of frequency and the 
wavevector $\vec{k}$, and is well behaved for weakly correlated systems such as band insulators or metals that behave as Landau Fermi 
liquids. Its real part shifts poles in the spectral function from the band energy $E(\vec{k})$, and its imaginary 
part broadens the poles, which are no longer Dirac deltas. (This corresponds to a finite quasiparticle life time due to electron-electron 
scattering.)  

The single site DMFT approximation, which assumes that the correlations are local to a single atomic 
site and works for a multitude of transition metal oxides, leads to a self energy $\Sigma(\omega)$ has no $\vec{k}$ 
dependence. $\Sigma(\omega)$ has a simple form in metals that behave as a Landau Fermi Liquid. Near the Fermi 
level, the imaginary part of $\Sigma$ goes to zero quadratically since the quasiparticles are long lived, and the real 
part of $\Sigma$ becomes a linear function of $\omega$. 
It is still possible to speak of a bandstructure with well defined bands, since the imaginary part of $\Sigma$ is zero, but 
the bandwidth is narrower than the DFT bandwidth by a factor of 
\begin{equation}
Z=\frac{1}{1-\frac{dRe(\Sigma)}{d\omega}}
\end{equation}
The reciprocal of $Z$ can be considered as a mass renormalization factor, since the electron effective mass calculated from DFT 
(often referred to as the \textit{band mass}) 
\begin{equation}
m_{\textrm{band}}^*=\hbar^2 \left( \frac{\partial^2 E(k)}{\partial k^2}\right)^{-1}
\end{equation}
is smaller than the mass approximated from the DMFT spectral function by a factor of $Z$
\begin{equation}
m_{\textrm{DMFT}}^* = Z^{-1} m_{\textrm{band}}^*
\end{equation}
A compound with no electronic correlations that cannot be captured by DFT at the LDA level has $Z=1$. Stronger electronic
correlations lead to a smaller $Z$, which approaches zero as one approaches the Mott insulating phase in the phase diagram. 

\begin{figure}[h]
\includegraphics[width=4.5in]{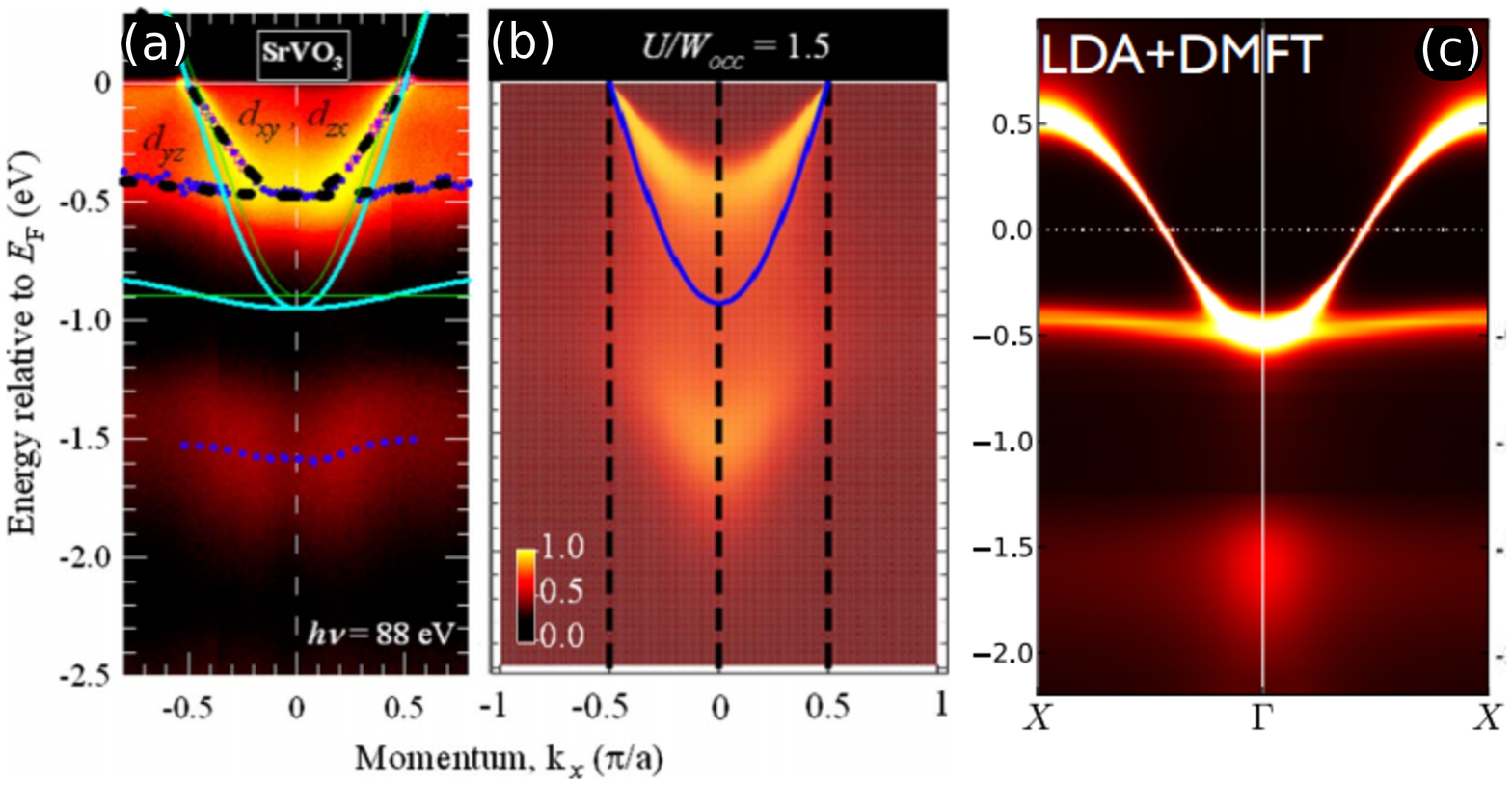}
\caption{ARPES and first principles results for the spectral function of correlated metal SrVO$_3$. (a) ARPES spectral weight from 
Ref. \cite{takizawa2009}. The dark blue dots are a fitted tight binding model, and the light blue lines are the DFT bands. (b) Spectral 
function from the one-band DMFT calculation from Ref. \cite{takizawa2009}. Dark blue line is the uncorrelated band structure used as the 
input to the DMFT calculation. (c) Results of the self consistent DFT+DMFT calculations from Ref. \cite{haule2015free}. Copyright (2009) 
and (2015) by the American Physical Society.}
\label{fig:SrVO3}
\end{figure}

The cubic perovskite oxide SrVO$_3$ \cite{dougier1975_srvo3} provides a clear illustration of this point. 
This compound has a simple band structure with only partially filled $t_{2g}$ bands of V crossing the Fermi level. 
Angle-Resolved Photoemission Spectroscopy (ARPES) measurements of Takizawa et al. \cite{takizawa2009} show 
that these $t_{2g}$ bands are relatively coherent (are not 
broadened very much by the imaginary part of the self energy), as reproduced in Fig. \ref{fig:SrVO3}a. 
The bandwidth 
predicted by DFT calculations, displayed as the blue lines superposed with the ARPES result, is larger by about a factor of 2. 

In order to correct the bandwidth, Takizawa et al. performed a DMFT calculation on a one-band model \cite{takizawa2009}. 
This model calculation considered a tight binding model that was obtained from the first principles DFT calculation, 
but it did not take into account all the bands present in the material.
Nevertheless, this calculation provides significant improvement over the DFT band structure. 
For the on-site interaction $U$ set to 60\% of the uncorrelated bandwidth $W$, $U=0.6 W$, Takizawa et al. 
found $Z\sim0.5$ which gives the correct 
bandwidth for the quasiparticle band (Fig. \ref{fig:SrVO3}b).
However, this one band approach overestimates the dispersion of the incoherent satellite at -1.5 eV that consists of the spectral weight 
transferred from the coherent band. This is possibly due to the omission of the other $t_{2g}$ bands that overlap \cite{takizawa2009}.  
LDA+DMFT calculations that don't omit the other $t_{2g}$ bands, such as the early calculations by Nekrasov et al. that employ a downfolding
scheme and considered 3 correlated $t_{2g}$ orbitals per V ion, also reproduce a similar $Z$ value \cite{nekrasov2006}. 
Relatively recently, a fully charge self consistent DFT+DMFT calculation performed by one of 
us (Birol) and Haule \cite{haule2015free} correctly reproduced not only the $Z$ factor of $Z\sim 0.5$, but also the weakly dispersive 
satellite at -1.5 eV (Fig. \ref{fig:SrVO3}c).

\begin{figure}[h]
\includegraphics[width=4.in]{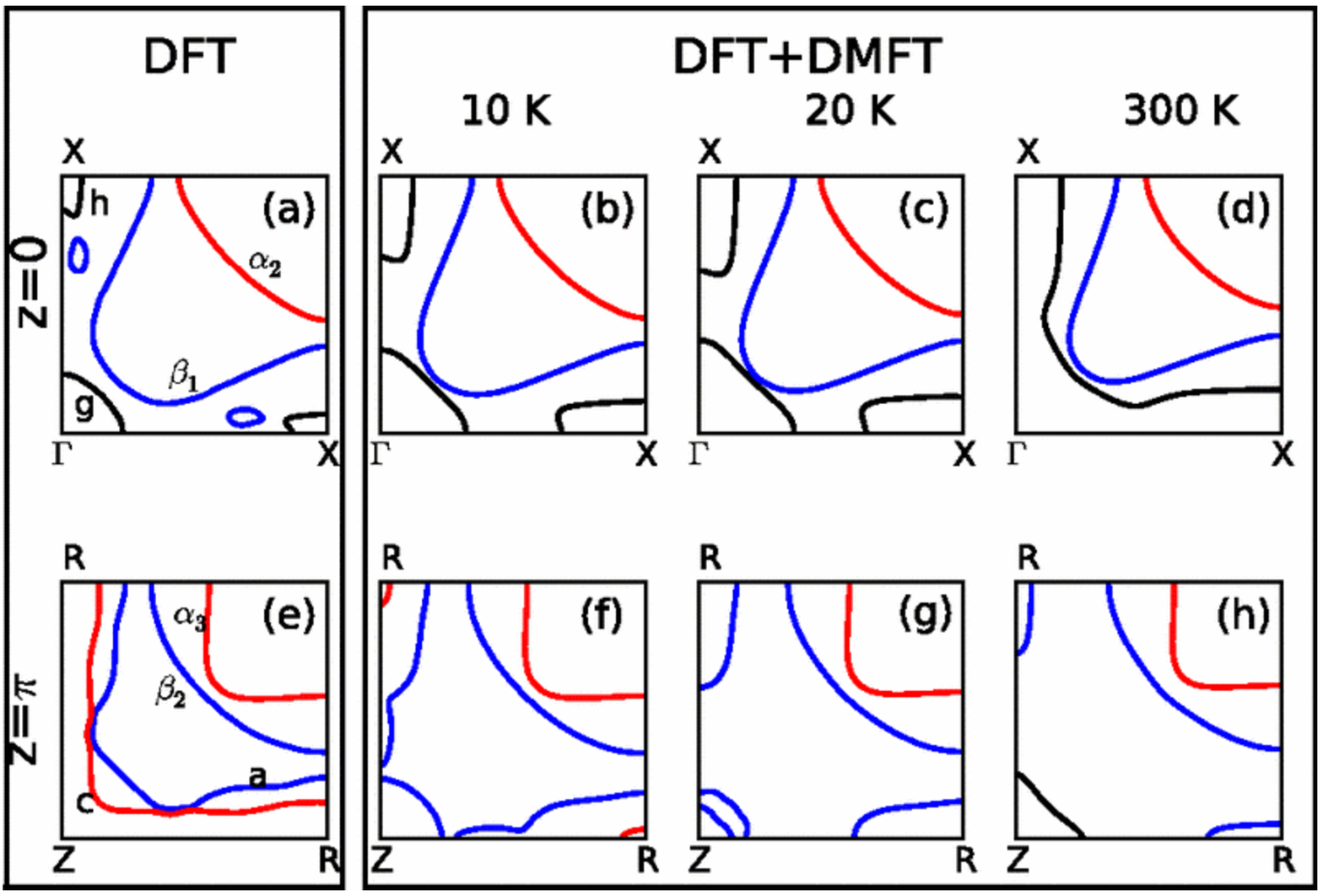}
\caption{Cuts through the Fermi surface of CeIrIn$_5$ at different $k_z$ values from DFT and DFT+DMFT. The DFT+DMFT 
approach allows performing the calculation at different temperatures to elucidate temperature evolution of the Fermi surface. 
Reproduced from Ref. \cite{choi2012}. Copyright (2012) by the American Physical Society.}
\label{fig:FSvsT_choi}
\end{figure}

The reason that one can easily define a band structure and observe sharp bands in SrVO$_3$ is that it is only a mildly correlated metal.
The heavy fermion compounds, by comparison, are extremely correlated Fermi liquid systems, and the effective mass of electronic 
quasiparticles in these compounds can be multiple orders of magnitude larger than in ordinary metals, or what DFT 
predicts for them \cite{hewson1997}. 
In these systems, the Fermi step is reduced from its uncorrelated value 1 so much that the observation of a Fermi surface and 
measuring a $Z$ (which is typically $\sim 0.01 - 0.001$) is very hard. An easy way to detect the signature of the very 
strong correlations in heavy fermion compounds is the anomalously large electronic specific heat Sommerfeld coefficient $\gamma$
in these systems 
\begin{equation}
C_{e}=\gamma T
\end{equation}
which is also renormalized with respect to its DFT value by the same amount as the effective mass \cite{fazekas1999}. 
\begin{equation}
\frac{\gamma}{\gamma_{band}}=\frac{m^*}{m_{band}^*}
\end{equation}
DFT+DMFT approach has been extensively used to study heavy fermion compounds as well. For example, LiV$_2$O$_4$, 
a frustrated, mixed valence spinel \cite{uehara2015} that exhibits the largest specific heat enhancement among the 
heavy fermion compounds that don't contain a rare earth ion \cite{kondo1997}, 
has been studied by Arita et al., who showed that there is a very sharp (heavy) quasiparticle peak near the 
Fermi surface \cite{arita1997}. 
Haule et al. \cite{haule2010, shim2007} studied the 115 heavy fermion materials CeIrIn$_5$, CeCoIn$_5$, and CeRhIn$_5$. 
Comparing the characters of the Ce $f$ electrons as obtained from DMFT shows that the localization tendency is highest 
in CeRhIn$_5$, and the electrons in the Iridium compound CeIrIn$_5$ display the most itinerant character; both in 
line with the experimental observations.
Later work by Choi et al. \cite{choi2012} showed that the electronic temperature directly affects the Fermi surface in these 115 
compounds. Even though DFT correctly reproduces the shape of the low temperature Fermi surface, it does not contain any temperature 
for the electrons. The DMFT calculation, on the other hand, naturally includes the electronic temperature and hence allows 
studying the evolution of the Fermi surface with temperature (Fig. \ref{fig:FSvsT_choi}).  

\section{CRYSTAL STRUCTURES AND LATTICE RESPONSE FROM DFT+DMFT}
\label{sec:structure}
\subsection{Error in lattice parameters of correlated materials from DFT}
DFT has evolved considerably since the original theorems of Kohn et al. \cite{hohenberg1964, kohn1965}. 
Part of this evolution was a transition from in-house codes written and maintained by small groups to 
large scale packages used by thousands of groups that increased precision and reproducibility \cite{lejaeghere2016}.  
Interestingly, the simplest approximation to the exchange-correlation energy, the so-called local density approximation (LDA) 
that was originally proposed by Kohn et al. in References \cite{hohenberg1964, kohn1965}, is still commonly used. 
LDA is surprisingly accurate in predicting crystal structures of band insulators and uncorrelated metals \cite{baroni2001}. 
It is well known to 
underestimate the lattice constant by $\sim 1\%$ because it does not assign an energy cost to a larger electronic density gradient, 
but there is no obvious reason why the error of such a simple approximation should be this small.  
Various generalized gradient approximations \cite{perdew1986} give even better results than LDA. For example, the PBEsol functional 
is developed specifically for solids \cite{PBE, PBEsol} and it often has an error of the order 
of few tenths of a percent for the lattice parameters \cite{haas2009}. 
Other details of the crystal structure (the positions of the atoms in 
the unit cell) and linear response properties such as phonon frequencies can also be precisely determined 
in band insulators using either LDA or its simple extensions \cite{baroni2001}. 
Historically, these methods provided significant levels of insight and 
quantitative accuracy in the study of crystallographic phase transitions, the best example being the ferroelectric transitions 
in oxides \cite{rabe2007, vanderbilt1998}. 

In certain compounds, such as the Mott-insulating 3d transition metal oxides, 
LDA often underestimates the lattice parameters with a much larger error margin. 
For example, performing a DFT calculation without magnetic ordering leads to 
an underestimation of the lattice constant of FeO by $\sim 7.7\%$ by LDA and $\sim 5.1\%$ by PBE (a type of GGA) 
compared to the experimental value in the paramagnetic state \cite{haule2015free}.
(PBE usually tends to \textit{over}estimate lattice constants in band insulators.) Performing the calculation with 
antiferromagnetic order reduces the error and leads to an underestimation of $\sim 3.6\%$ and $\sim 0.7\%$ by 
LDA and PBE respectively. While this is a smaller error, it is nevertheless 
significantly higher than that in the results obtained for band insulators, pointing to the presence of a physical 
reason that leads to enhanced overbinding of the lattice in compounds like FeO. 

\begin{figure}[h]
\includegraphics[width=4.0in]{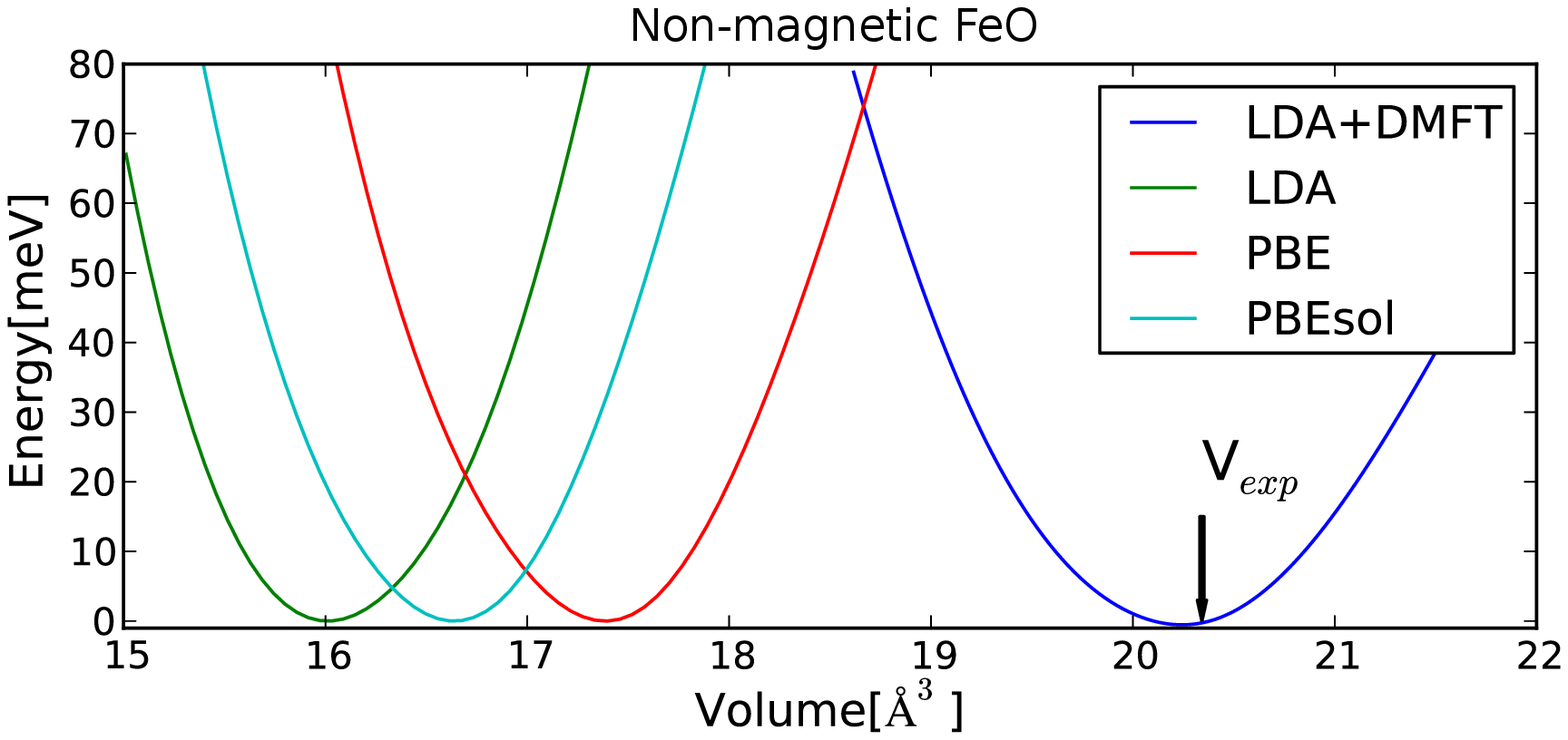}
\caption{Energy as a function of unit cell volume for cubic FeO without magnetic order, calculated using DFT with LDA (green), 
PBE (red) and PBEsol (light blue) exhange correlation functionals, as well as DFT+DMFT (dark blue). The minimum of the energy 
gives the predicted volume. The black arrow denotes the experimentally observed value $V_{exp}$. Reproduced
from Ref. \cite{haule2015free}. Copyright (2015) by the American Physical Society.}
\label{fig:FeO}
\end{figure}

\begin{figure}[h]
\includegraphics[width=4.0in]{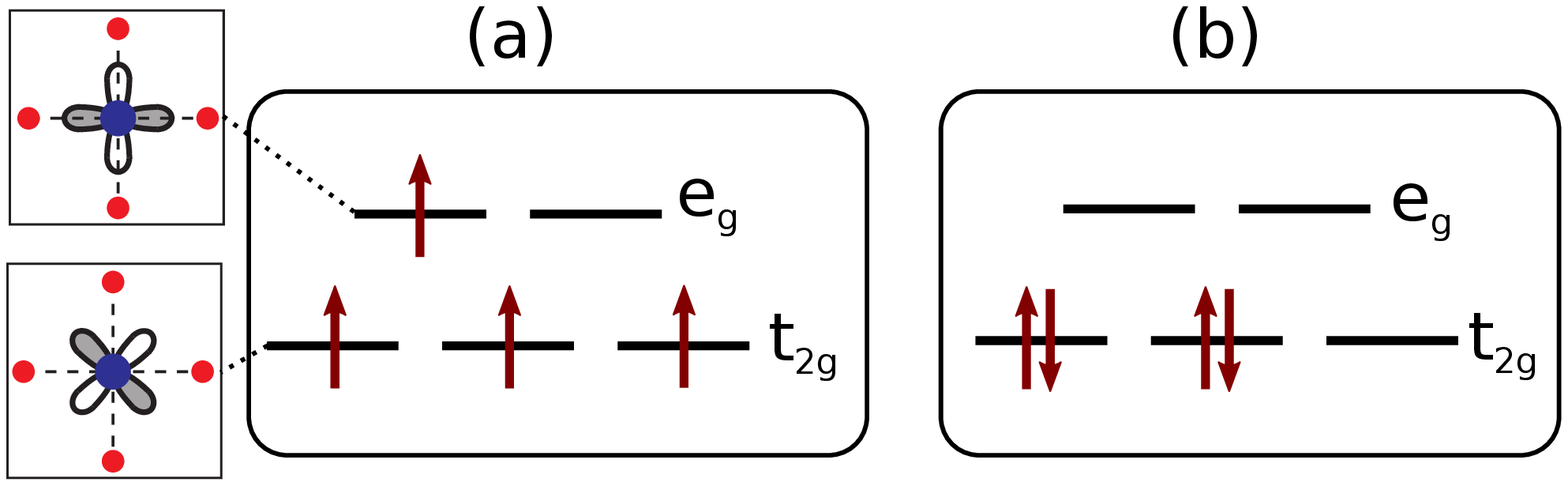}
\caption{Sketch of the atomic energy levels and the electronic configuration of a transition metal with 4 electrons on its $d$ shell, 
under a cubic crystal field. (a) The high-spin configuration, where the higher $e_g$ states, which have lobes extended towards 
the oxygen anions, are partially occupied. (b) The low-spin configuration, where only the lower $t_{2g}$ orbitals without lobes 
extended towards oxygens are occupied.}
\label{fig:t2geg}
\end{figure}

A major reason for the underestimation of the lattice constant, especially in nonmagnetic DFT calculations, is the 
strong underestimation of the local magnetic moments by DFT at the LDA or GGA level. The paramagnetic state 
often involves large local, atomic magnetic moments. For example, in a transition metal cation with 4 valence 
electrons under a cubic crystal field, all of the electrons often have parallel spin due to 
atomic Hund's coupling (Fig. \ref{fig:t2geg}a) even in the paramagnetic phase. 
In the absence of magnetic order, the atomic magnetic moment is fluctuating, and $\langle S_z \rangle =0$, even 
though $\langle S_z^2 \rangle \neq 0$. Kohn-Sham DFT cannot directly capture this phase, and a simple DFT calculation 
without magnetic ordering simulates a state where  $\langle S_z \rangle =0$, and $\langle S_z^2 \rangle = 0$. This 
often corresponds to the low-spin configuration in Fig. \ref{fig:t2geg}b. 

The underestimation of the lattice constants by LDA/GGA can to a large extent be explained by the absence of a local, 
fluctuating magnetic moment. 
In an oxide like FeO, the bonding of a high-spin cation with an electron in the higher lying $e_g$ orbitals (Fig. \ref{fig:t2geg}a) is 
in general very different from that of a low-spin cation without any electrons in the $e_g$ shell. This is because of the 
fact that $e_g$ orbitals are extended towards the oxygen anions, and are $\sigma$ bonding with them \cite{khomskii2014}. 
Electrons on the $e_g$ orbitals strongly repel the oxygen anions, and hence favor a larger lattice constant. A nonmagnetic DFT 
calculation that places all the electrons to the lower lying $t_{2g}$ orbital strongly favors an underestimated lattice constant. 
In addition to the large errors in lattice parameters, the phonon spectra that DFT predicts for these compounds are 
often both quantitatively and qualitatively wrong. 
Imposing a magnetic order in LDA fixes part of the error, but is often not enough to fix all of it,
since obtaining a realistically large ordered moment in transition metal oxides often requires the correction of the on-site 
Coulomb interaction by a Hubbard-U term as well \cite{DFTU:LDAUTYPE1_1, DFTU:Dudarev, DFTU:LDAUTYPE1_2}. 
Also, in many compounds such as in the stripe-type antiferromagnetically ordered iron pnictides, the antiferromagnetic order 
breaks lattice symmetries in addition to the time reversal symmetry \cite{pnictide_stripe}. In these systems, the 
antiferromagnetic state does not provide a good approximation to the paramagnetic phase for calculating crystal structure properties. 
DFT+DMFT approaches, on the other hand, bring the capability to perform calculations in a truly PM phase, with 
nonzero local fluctuating moments, as well as ordered moments. This, in addition to the dynamical correlations that 
DMFT introduces, has been recognized early on as a means to correctly reproduce the lattice parameters and phonon 
spectra of correlated materials from first principles. 
The result of the calculation of lattice parameter of paramagnetic 
FeO by DMFT, displayed in Fig. \ref{fig:FeO}, exemplifies this point. 

\subsection{Phonons and structural stability of elemental Iron}
Early examples of work on the structural stability and response of correlated materials 
include the study by Savrasov et al. \cite{savrasov2003} on the phonon 
band structure of MnO and NiO, and the work of Dai et al. \cite{dai2003} which reproduced not only the phonon 
spectra but also explained the unusually large anisotropy in the elastic properties of this compound \cite{dai2003}. 
Both of these studies employed the Hubbard-I approximation \cite{hubbard1963} to solve the DMFT impurity problem. 
(The Hubbard-I approximation does not give as precise results as the state of the art Monte Carlo approaches, but it is 
computationally much cheaper and does not require the relatively delicate analytical continuation step.) 
DFT+DMFT approach also made important contributions to applied problems, for example by 
explaining the low thermal conductivity of nuclear fuel materials UO$_2$ and PuO$_2$ \cite{yin2008}.  

Even one of the oldest materials known to mankind, elemental iron, is not exempt from electronic correlation 
effects. It was recognized early on that even though elemental metals have seemingly large 
bandwidths, the 3d transition metals also have large 
on-site Coulomb interactions, and hence their electronic and magnetic properties are more correctly 
given by a DFT+DMFT treatment than by DFT alone \cite{katsnelson2000, lichtenstein2001, kotliar2006, yang2001}.  
As a function of temperature, elemental iron undergoes multiple phase transitions before melting. Fe is a 
ferromagnet with a body-centered cubic (BCC) crystal structure ($\alpha$-phase) at low temperature. The crystal 
structure becomes face-centered cubic (FCC) at $\sim 1185$ K, just $\sim 140$ K above Curie temperature 
at which the ferromagnetic order disappears ($\gamma$-phase); but it becomes BCC again at $\sim 1670$ K 
($\delta$-phase), close to its melting temperature. 
Nonmagnetic DFT calculations predict strong lattice instabilities (imaginary phonon frequencies) for both the 
$\alpha$ and the $\gamma$ structures, which poses a clear contradiction with the experimental observations. 
Leonov et al. employed a Wannier based DFT+DMFT approach and the Hirsch-Fye algorithm for the impurity 
solver \cite{hirsch1986} to address the structural stability of Fe near its 
phase transitions \cite{leonov2011, leonov2012, leonov2014}, and reached the conclusion that ``electronic 
correlations determine the phase stability of iron up to the melting temperature'' \cite{leonov2014}. 

\begin{figure}[h]
\includegraphics[width=3.5in]{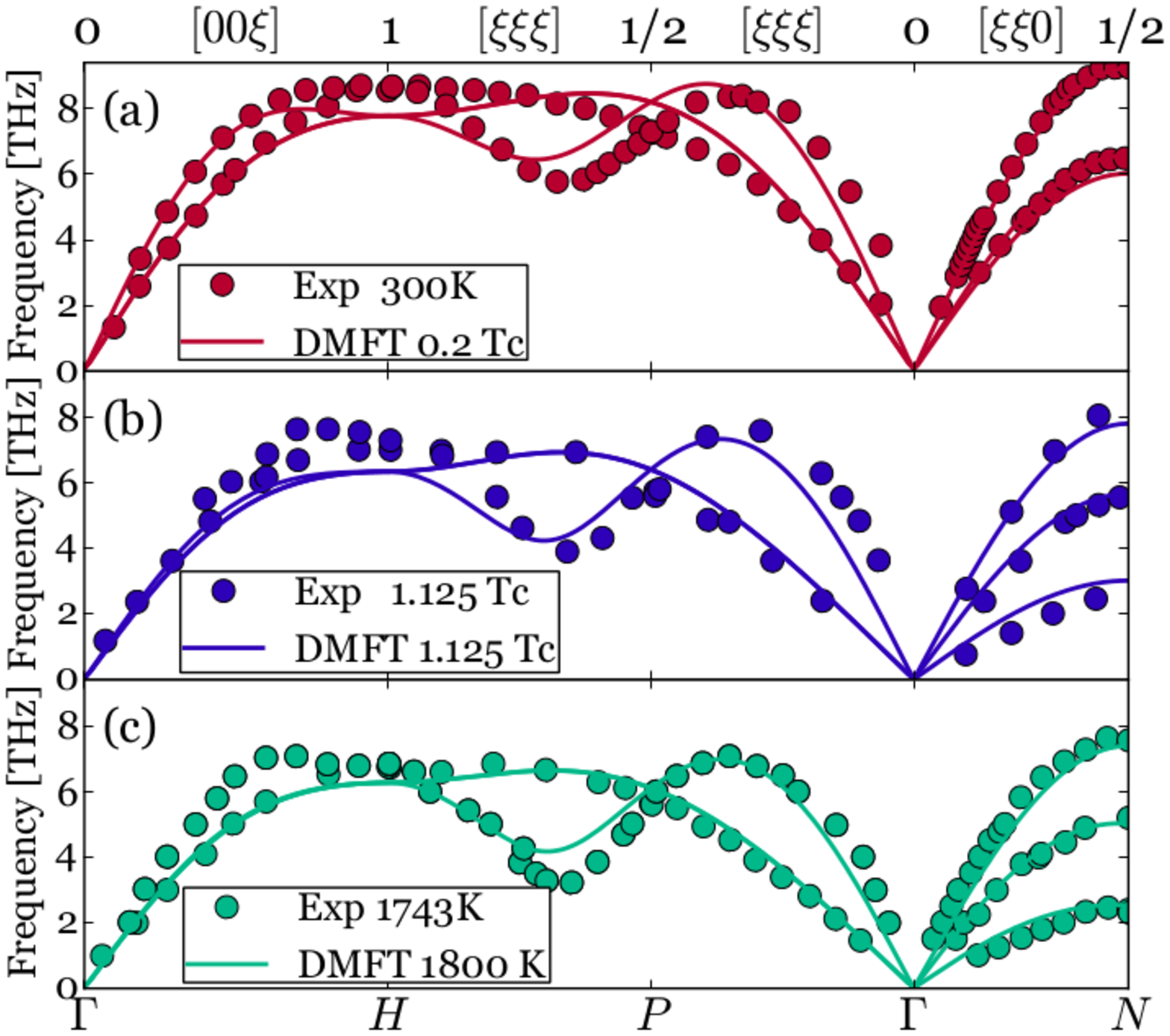}
\caption{Phonon dispersions of elemental iron (a) at the ferromagnetic $\alpha$ phase, (b) at the 
$\alpha$ phase with no magnetic order at higher temperature, and (c) at the $\delta$ phase. Reproduced
from Ref. \cite{han2018}. Copyright (2018) by the American Physical Society.}
\label{fig:fe_disp}
\end{figure}

In Fig. \ref{fig:fe_disp}, we present the phonon dispersions of iron at different phases and temperatures, 
reproduced from Ref. \cite{han2018} by Han, Birol, and Haule. The DFT+DMFT approach that we used in this work 
is fully charge self consistent, and uses the numerically exact Continuous Time Quantum Monte Carlo (CTQMC) impurity solver 
\cite{haule2007, haule2010}. 
In order to overcome the Monte-Carlo numerical noise in the calculations, in addition to good statistics (which comes 
at the cost of high computational cost), a stationary implementation of DFT+DMFT is necessary. The stationary implementation 
we used \cite{haule2015free} allows the calculation of the free energy (including the electronic entropy) and accurate 
forces on the atoms \cite{haule2016, haule2018}. This implementation of the forces also takes into account the electronic entropy, 
which can be particularly important near a Mott or spin state transition. 
(For example, LaCoO$_3$ displays both a spin-state transition and anomalous thermal expansion \cite{thornton1986}; 
and the high temperature crystal structure is energetically favorable only when the electronic entropy 
is taken into account in DFT+DMFT \cite{chakrabarti2017}.)

\begin{figure}[h]
\includegraphics[width=4.0in]{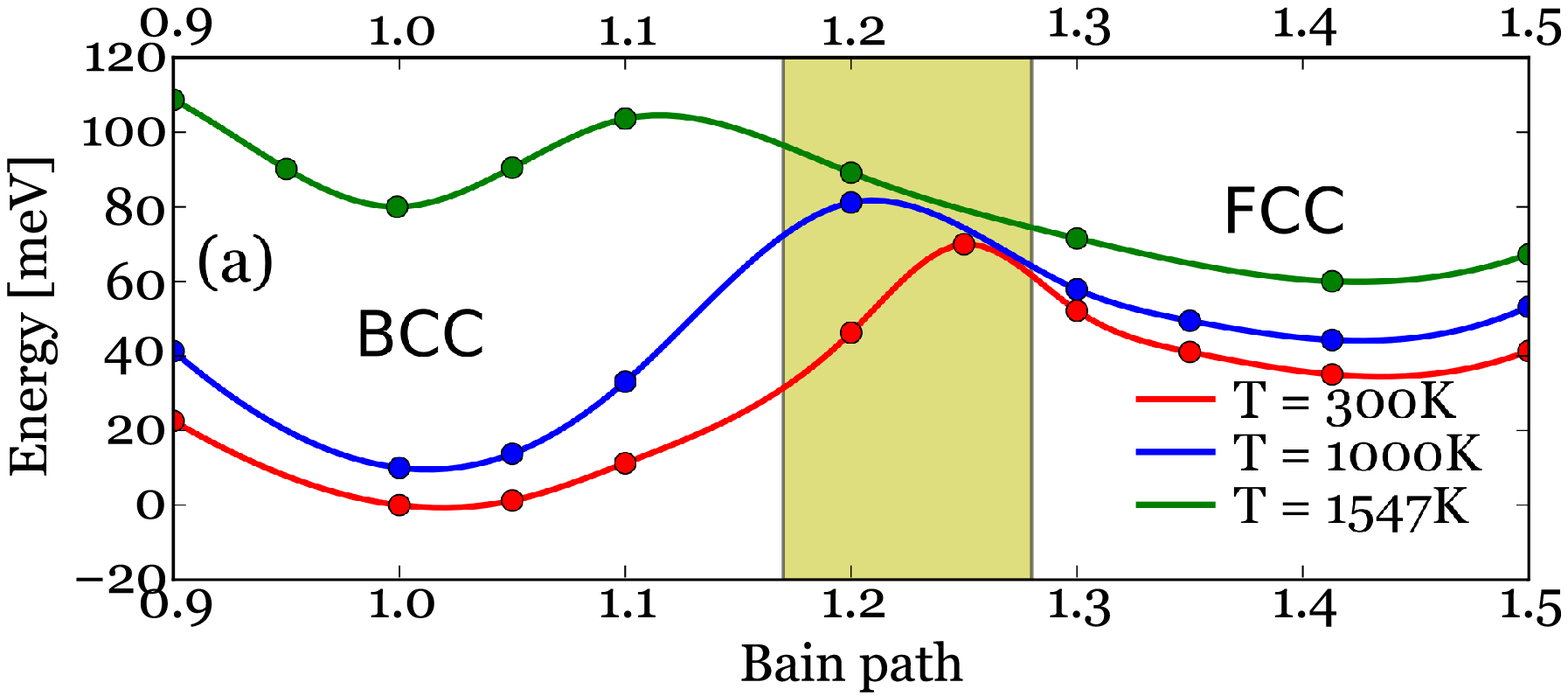}
\caption{The free energy of elemental iron along the Bain path as obtained from DFT+DMFT at different temperatures. Reproduced
from Ref. \cite{han2018}. Copyright (2018) by the American Physical Society.}
\label{fig:fe_bain}
\end{figure}

Our results show that iron is dynamically stable at all temperatures in all of its phases; in other words, 
the fully charge self-consistent DFT+DMFT calculations do not predict any unstable phonons at any temperature or structure. 
This result explains the reason behind the phonon softening observed near T$_C$ \cite{satija1985} as merely 
the melting of the magnetic order, and not the proximity to the $\alpha \rightarrow \gamma$ structural transition. 
The total energy along the Bain path (the path that involves both strain and ionic displacements, and connects the 
BCC and FCC structure \cite{bain1924}) has two local minima corresponding to FCC and BCC at all temperatures, but the relative 
energy of these minima change as the system crosses the structural transition temperature (Fig. \ref{fig:fe_bain}). 
This is a surprising result that goes against the common assumption that the softening observed near the magnetic 
transition is a precursor of the martensitic transition \cite{neuhaus1997}, and demonstrates the power of the DFT+DMFT 
approach in providing insight on the coupling between crystal structure and correlated electronic phases. 

While it is not possible to simulate a paramagnetic phase directly within DFT+LDA, 
it is possible to approximate certain properties of materials in the paramagnetic phase using only DFT+U by employing a multi-step 
approach and constructing multiple large supercells 
with different magnetic configurations. For example, the \textit{special quasirandom structures}, which originally 
emerged from studies of alloys \cite{zunger1990}, were recently used to study MnO, FeO, CoO, and NiO by Trimarchi 
et al. \cite{trimarchi2018}. 
A similar method was also used by K\"ormann et al. to calculate the phonon spectra of paramagnetic iron \cite{kormann2012}, 
which lead to studies of pressure dependence of phonons, and magnon-phonon coupling in this system \cite{ikeda2014, kormann2014}.
Other groups employed a similar approach to study the temperature dependence of phonons in Earth's core conditions \cite{luo2010} 
and near the $\gamma$-$\delta$ phase transition \cite{lian2015} with the help of an auxiliary Heisenberg magnetic model 
used to simulate the magnetic state at nonzero temperature. 
While all of this work provides valuable insight, the idea of replacing a fluctuating moment in a paramagnet with a 
spatial disorder of magnetic moments, which relies on ergodicity \cite{han2018}, falls short of simulating the 
dynamic fluctuations at finite temperature. 

Other examples of DFT+DMFT studies on the coupling between the electronic correlations, and the crystal structure and response of correlated 
materials include the work of Kune\v{s} et al. on the pressure induced volume collapse and metallization of MnO \cite{kunes2008}, 
the Jahn-Teller effect in KCuF$_3$ \cite{leonov2008}, and the extensive work on rare earth nickelates, including the explanation 
of the site selective Mott transition in these compounds \cite{park2012, haule2017}. These last two examples point out to 
the success of DFT+DMFT in studying structural phase transitions in correlated materials.

\section{NONLOCAL CORRELATIONS}
\label{sec:nonlocal}
\subsection{DMFT with other extensions of DFT+LDA}

While single site DMFT corrects the shortcomings of DFT+LDA by adding dynamical correlations that are local to an atom, 
it is not always sufficient for correctly reproducing the properties of 
materials where nonlocal correlations (either static or dynamic) beyond LDA are important. 
An example of such a compound is Cerium(III) Oxide Ce$_2$O$_3$ \cite{ce2o3_exp}.
Hybrid functionals that are now commonly used in DFT studies include the nonlocal Fock exchange 
\cite{hyb1,hyb2,hyb3,hyb4,hyb5,hyb6,hyb7} and correctly predict Ce$_2$O$_3$ to be an insulator \cite{ce2o3_hybrid}. 
They additionally improve lattice properties significantly with respect to LDA \cite{ce2o3_hybrid}, but 
they do not reproduce the dynamically correlated 4f Hubbard bands correctly in the paramagnetic state.
A natural approach to bring together the best of both hybrid functional and DMFT methods is taken by Jacob et al. 
\cite{jacob2008}, who performed a DFT + Hybrid Functional + DMFT calculation on Ce$_2$O$_3$. In this approach, the 
exact exchange included in the hybrid functional corrects the magnitude of the Cerium $d$ - Oxygen $p$ gap, and the single site 
DMFT corrects the dynamically correlated Cerium f states.
%

Another approach, which can be used to take into account the screening by long-range Coulomb interactions, involves 
combining GW with DMFT. The GW approach is known to produce impressive results in semiconductors \cite{giovanni2002}, 
and attempts to interface it with DMFT were undertaken as early as early 2000s \cite{kotliar2001, biermann2003}. 
Modern applications of GW+DMFT to the correlated metal SrVO$_3$ show that while the dynamical renormalizations are essentially 
local to the vanadium ion in this compound, the nonlocal correlations screen the Fock exchange and dramatically modify  
the unoccupied states \cite{tomczak2012, tomczak2014}. Some correction to the position of the lower Hubbard band is also
reported in SrVO$_3$ \cite{taranto2013}. 
On a completely different type of materials, 
Hansmann et al. applied the GW+DMFT to effective Hamiltonians obtained from first principles 
calculations of Si(111) surface with adatoms such as Sn, Si, C, and Pb, and verified the expectation that the 
nonlocal effects are particularly important in charge density wave systems \cite{hansmann2013}.  
Another recent development in GW+DMFT is the implementation of first principles quasiparticle self-consistent GW + DMFT 
(QSGW+DMFT) \cite{choi2016}. 

\subsection{Cluster DMFT}
\begin{figure}[h]
\includegraphics[width=4.in]{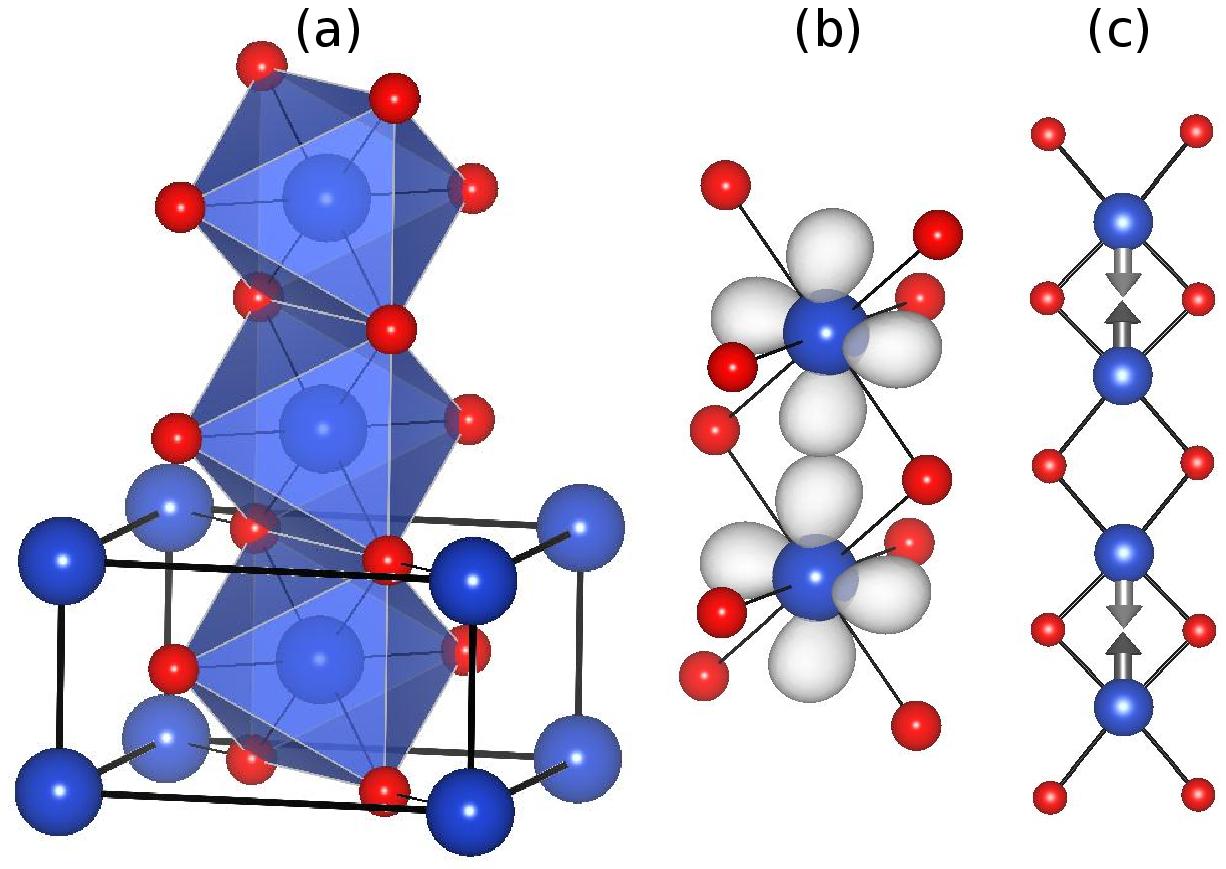}
\caption{(a) The Rutile crystal structure of VO$_2$ at high temperature. The tetragonal crystallographic unit cell consists of 
two formula units, with two V ions on the corner and the body center of the cell. The oxygen octahedra surrounding the V cations 
form edge-sharing chains along the crystallographic c axis. (b) One of the $t_{2g}$ orbitals on each V ion has lobes 
extended along these octahedral chains, enabling significant direct V-V hopping of electrons. (c) Below 340 K, V ions along 
each chain dimerize and lower the symmetry to monoclinic.}
\label{fig:vo2}
\end{figure}

It is possible to explicitly prove that nonlocal dynamic correlations are not important for a particular class of compounds. 
For example, Semon et al. \cite{semon2017} considered a model that represents the iron pnictide and chalcogenide superconductors 
and showed that it is justified to use single-site DMFT for these compounds. 
On the other hand, there are compounds such as 
VO$_2$, where nonlocal dynamic correlations are essential. VO$_2$ is a metal and 
has the tetragonal rutile structure above $\sim 340$ K (Fig. \ref{fig:vo2}a). In this structure, the vanadium cation is 
in the center of an oxygen octahedron. The octahedra are corner sharing in two dimensions, but are edge sharing along the 
crystallographic c axis. These edge sharing octahedral chains lead to a smaller distance between the nearest neighbor cations, which results 
in direct V-V interactions, studied in detail for many decades \cite{goodenough1971}. 
One of the $t_{2g}$ orbitals of the V cation has lobes pointing along the direction of the edge-sharing octahedral chains. 
This orbital has overlap with the nearest neighbor V cations (Fig. \ref{fig:vo2}b), and is responsible for coincident 
metal-insulator and monoclinic dimerization transitions observed at 340 K (Fig. \ref{fig:vo2}c). The nature of this transition, 
in particular whether the low temperature phase is a Peiers or a Mott insulator has been the subject of debate. 

Both DFT \cite{eyert2002} and DFT+(single site)DMFT calculations predict a metallic phase in the monoclinic 
structure \cite{weber2012}, which is not in line with the experimental observations. 
In order to take into account the nonlocal correlations, Biermann et al. \cite{biermann2005} 
performed cluster DMFT (C-DMFT) calculations on the Wannier functions obtained from DFT. In C-DMFT, the impurity 
is considered to consist of more than one V atoms, and hence correlations that are not local to an atom can also be included in the 
DMFT self energy. While this approach increases the computational cost significantly (due to multiple reasons 
including the larger number of orbitals in the impurity), it is necessary for taking into account dynamical nature of 
the nonlocal correlations. Biermann et al.'s calculations reproduced the insulating monoclinic phase, and showed that 
``dynamical V-V singlet pairs due to strong Coulomb correlations is necessary'' for the formation of the Peierls gap in VO$_2$. 
Lazarovitz et al.'s similar calculations on a downfolded model addressed the effect of strain on the monoclinic transition 
\cite{lazarovits2010}. Recent first principles, self consistent DFT+DMFT calculations \cite{brito2016, brito2017} support this 
picture, and emphasize the importance of nonlocal correlations both in VO$_2$, and its less correlated cousin NbO$_2$. 

\section{SUMMARY \& OUTLOOK}
\label{sec:summary}
2019 marks 20 years since Franceschetti and Zunger introduced the \textit{inverse bandstructure problem} of finding out 
what compound gives rise to a desired functionality \cite{franceschetti1999}. In the two decades since, first principles DFT has been 
extremely successful in not just explaining and supporting experimental observations, but also providing predictions 
and guiding the experiments in new directions and to new materials. 

This procedure, often dubbed \textit{materials by 
design}, is however limited by the theoretical tools available, in particular, the shortcomings of DFT. 
The method reviewed in this article, first principles DFT+DMFT, is a leading tool that can to a good extent correct the 
errors of DFT when applied to correlated materials. Recent developments in the methodology, and implementations of 
DMFT are now enabling a larger community to work on new problems, and come up with verifiable predictions. 
These developments have finally turned \textit{correlated materials by design} into reality, and more insight and predictions 
from DFT+DMFT are sure to follow \cite{adler2018}. 

\section*{DISCLOSURE STATEMENT}
The authors are not aware of any affiliations, memberships, funding, or financial holdings that
might be perceived as affecting the objectivity of this review.

\section*{ACKNOWLEDGMENTS}
This work was supported by the NSF-DMR under grant DMREF-1629260.
We acknowledge useful discussions with Onur Erten.

\end{document}